\documentstyle{article}
\begin{document}
\title{A model of gravitation with global U(1)-symmetry}
\author
{By M.A.Ivanov \\ Chair of Physics, \\ Belarus State
University of Informatics and Radioelectronics, \\ 6 Pyatrus
Brovka Street, 220027, Minsk, Republic of Belarus.
\\ E-mail: ivanovma@gw.bsuir.unibel.by}
\maketitle
   \begin{abstract}
     It is shown that an embedding of the general relativity $4-$space into
a flat $12-$space gives a model of gravitation with the global
$U(1)-$symmetry and the discrete $D_{1}-$one. The last one may be transformed
into the $SU(2)-$symmetry of the unified model, and the demand of
independence of $U(1)-$ and $SU(2)-$transformations leads to the
estimate $\sin^{2}\theta_{min}=0,20$ where $\theta_{min}$ is an analog of
the Weinberg angle of the standard model.
\end{abstract}
\par

\section[1]{Introduction }

In Einstein's theory of gravitation, the flat Minkowski space is not
used as a background one that differs the theory from other physical
theories. From a geometrical point of view, a curved space can be embedded
into a flat one of an enlarged dimension. A similar embedding, but into a
nonflat space with the additional spinor coordinates, is factically used
in the Ashtekar approach \cite{A}, that makes possible to introduce the new
variables for a description of gravitational field.
\par Many-dimensional spaces are widely used in physics, for example,
in theories of supersymmetry, supergravitation, and
superstrings \cite{G}. In theories of the Kaluza-Klein type in the 8-space,
a description of a system of generations of the fundamental fermions is
possible, by which primary postulates of the standard model are the
consequences of the hypothesis about compositness of the fermions \cite{I}.
     It is shown in this paper, how one can realize such an embedding of
the general relativity 4-space into a flat 12-space. The additional
coordinates are choosen to have a clear interpretation. If in the
four-dimensional case, three points --- a point of the Minkowski space, a
trial material point (its position), and some point of observation --- can be
described by one set of the coordinates $x$, then in the model it is
postulated, that these three points are described by three independent sets
of coordinates $x, A, B$ accordingly. \par
     At an initial stage, the connection is introduced with a trivial
curvature in the model, which is caused by local rotations under
transitions from one four-dimensional subspace to another. A linear
deformation of such the connection lets to introduce the 4-manifold with
a non-trivial curvature. Then one obliges the Ricci tensor of the
deformed connection to Einstein's equation. The equations, which picks
out the curved manifold, will be the algebraic ones for components of the
connection of flat space. All three flat subspaces are locally
isomorphic to each other, i.e. the additional coordinates have the
"vector" kind. \par
     The model lets a natural unification with the composite fermions model
by the author \cite{I}, for which one needs to enlarge the space dimension
up to 16. It is an important fact, that the model has the $U(1)-$symmetry,
therewith it is the symmetry of field equations. Another interesting fact
is an existence of  the discrete $D_{1}-$symmetry. By an introduction of
spinor fields of the composite fermions with a demand of conservation its
norm \cite{I}, the last symmetry will lead to the global $SU(2)-$symmetry
of the model. But the parameters of the $U(1)-$transformations and the
$SU(2)-$ones will depend from each other in a general case. To provide its
independence, both conditions are necessary: 1) the $SU(2)-$doublet states
are massless, and 2) a rotation must be executed on some angle in the
parameter space. It is
similar to a situation with the Weinberg angle rotation in the standard
model, still a logical sequence of actions differs from the one that leads
to an introduction of the Weinberg angle.

\section[2]{Connections with a trivial curvature, introduced by mappings
in a flat 12-space}

\par
     Let us consider the flat 12-space $(x^{a}, A^{\tilde \mu}, B^{\mu}),$
where $(x)$ is the Minkowski space, $(A)$ is a trial body (a material
point) coordinate space, and $(B)$ is an observation point coordinate
space. Let us use simbols $a, b, \ldots \;$ for indices in $(x)-$subspace,
  $\tilde \alpha, \tilde \beta, \ldots \;$ in $(A)-$one, and $\alpha,
\beta \ldots \;$ in $(B)-$one. I.e. we shall suggest that the
diffeomorphisms exist: $x^{a} \to A^{\tilde \mu}(x), x^{a} \to B^{\mu}(x),$
which describe motions of a trial body and of an observation point in the
flat space $(x)$ in some systems of reference. These mappings are
characterized by the functions $h^{\tilde \mu}_{a} \equiv \partial
A^{\tilde \mu} / \partial x^{a}$ and $h^{\mu}_{a} \equiv \partial
B^{\mu} / \partial x^{a},$ for which the reverse functions exist:
$h^{\mu}_{a} h_{\mu}^{b} = \delta^{b}_{a},$ e.c. Let us consider that the
functions $h^{\tilde \mu}_{a}$ and $h^{\mu}_{a}$ describe the local
Lorentz transformations  $dx^{a} \to dA^{\tilde \mu}(x), dx^{a} \to dB^
{\mu}(x),$ if one projects both flat 4-spaces $(A)$ and $(B)$ onto $(x).$
\par
     By the sequential mappings $$x^{a} \to A^{\tilde \mu} \to B^{\mu}(x),$$
one has for the metric tensors of the subspaces $(x), (A), (B):$
$$\eta ^{ab} \to \eta^{\tilde \mu \tilde \nu} \to \eta^{\mu \nu}$$
and for the corresponding metric connections:
$$0=  \Gamma^{a}_{bc} \to \tilde \Gamma^{\tilde \mu}_{\tilde \nu \tilde
\epsilon} \to \Gamma^{\mu}_{\nu \epsilon},$$
where \cite{L}  $(h_{\tilde \mu}^{\mu} \equiv \partial B^{\mu} / \partial
A^{\tilde \mu}):$
\begin{equation}
\Gamma^{\mu}_{\nu \epsilon} = h_{\tilde \mu}^{\mu} h^{\tilde \nu}_{\nu}
h^{\tilde \epsilon}_{\epsilon}  \tilde \Gamma^{\tilde \mu}_{\tilde \nu
\tilde \epsilon} + h^{\mu}_{\tilde s} \partial_{\nu} h^{\tilde s}
_{\epsilon},
\end{equation}
\begin{equation}
 \tilde \Gamma^{\tilde \mu}_{\tilde \nu \tilde \epsilon} = h^{\tilde
\mu}_{a} \partial_{\tilde \nu} h^{a}_{\tilde \epsilon}.
\end{equation}
Let us denote the first part of (1) as $\tilde \Gamma^{\mu}_{\nu \epsilon}$
and the second one as $\bar \Gamma^{\mu}_{\nu \epsilon}$ and rewrite (1) as
\begin{equation}
 \Gamma^{\mu}_{\nu \epsilon} = \tilde \Gamma^{\mu}_{\nu \epsilon} + \bar
\Gamma^{\mu}_{\nu \epsilon}.
\end{equation} \par
     Relatively to the local coordinates $dB^{\mu},$ i.e. for $\partial_{\mu} =
\partial / \partial B^{\mu}$ in the definition of the curvature tensor:
\begin{equation}
R^{\alpha}_{\beta \gamma \delta} \equiv 2(\partial_{[ \gamma} \Gamma^
{\alpha}_{\delta ] \beta} +
\Gamma^{\alpha}_{\epsilon [ \gamma}  \Gamma^{\epsilon}_{\delta ] \beta}),
\end{equation}
the curvature of the connections $\Gamma^{\mu}_{\nu \epsilon}$ and $\bar
\Gamma^{\mu}_{\nu \epsilon}$ is equal to zero. The incomplete connection
$\tilde \Gamma^{\mu}_{\nu \epsilon}$ (the tensor part of $ \Gamma)$ has
a nontrivial curvature, but it is equal to zero by $\tilde \Gamma^{\mu}_{\nu
\epsilon} = 0,$ therefore the one cannot be used as Einstein's connection.

\section[3]{A linear deformation of the connection}

\par
     The connection $\gamma^{\mu}_{\nu \epsilon},$ which can be Einstein's
one on some 4-manifold, should have the following properties: 1) it must
sutisfy the transformation law (1) under transition to a new system of
reference of an observer $B^{\mu} \to C^{\mu};$ 2) its curvature tensor
must be not trivial: $r^{\alpha}_{\beta \gamma \delta} \not\equiv 0;$ 3) $
r^{\alpha}_{\beta \gamma \delta} \not\equiv 0,$ if  $\gamma^{\mu}_{\nu
\epsilon} = 0.$ \par
     The linear form $\gamma$ of the connections $\tilde \Gamma$ and $\bar
\Gamma,$ with one parameter $f,$ satisfies these demands:
\begin{equation}
\gamma^{\mu}_{\nu \epsilon} = f \tilde \Gamma^{\mu}_{\nu \epsilon} + \bar
\Gamma^{\mu}_{\nu \epsilon}.
\end{equation}
The transformation $\Gamma = \tilde \Gamma + \bar\Gamma \to \gamma = f
\tilde \Gamma + \bar \Gamma$ is called here a linear deformation of the
connection. In the paper, the parameter $f$ is global that provides the
global $U(1)-$symmetry of the model (with the peculiarity which is
discussed below). \par
     The curvature tensor of this connection relatively to the local
coordinates $dB^{\mu}$ is nontrivial by $f \not= 0; 1:$
\begin{equation}
r^{\alpha}_{\beta \gamma \delta} = 2(f^{2} - f) \tilde \Gamma^{\alpha}_
{\epsilon [  \gamma} \tilde \Gamma^{\epsilon}_{\delta ]  \beta},
\end{equation}
if $\tilde \Gamma^{\mu}_{\nu \epsilon} \not\equiv 0.$
Under the condition $\gamma^{\mu}_{\nu \epsilon} =0, $ we have $$
r^{\alpha}_{\beta \gamma \delta} = ((f - 1)/2) \tilde R^{\alpha}_
{\beta \gamma \delta},$$ where $\tilde R^{\alpha}_{\beta \gamma \delta}$
is the curvature tensor of the incomplete connection $\tilde \Gamma.$

\section[4]{Picking out of the four-dimensional curved manifold}

\par
     Let us denote as $g_{\mu \nu}$ the metric tensor which corresponds
to the metric connection
$ \gamma^{\mu}_{\nu \epsilon}.$  In the flat 12-space, let us pick out the
four-dimensional curved manifold $\Sigma^{4}$ with the metric tensor
$g_{\mu \nu}$ and the metric connection $ \gamma^{\mu}_{\nu \epsilon},$
on which Einstein's field equations are satisfied:
\begin{equation}
r_{\mu \nu} = k (T_{\mu \nu} - g_{\mu \nu} T/2),
\end{equation}
where $ r_{\mu \nu} $ is the Ricci tensor, $k$ is Einstein's constant,
and $T_{\mu \nu}$ is the matter energy-momentum tensor. \par
     We have
\begin{equation}
r_{\mu \nu} = 2(f^{2} - f) \tilde \Gamma^{\alpha}_{\epsilon [  \alpha}
\tilde \Gamma^{\epsilon}_{\nu ]  \mu}.
\end{equation}
Then by $T_{\mu \nu} =0,$ the manifold $\Sigma^{4}$ are picked out
by the algebraic equations for $\tilde
\Gamma^{\mu}_{\nu \epsilon},$ which do not depend on the parameter $f$
(under the condition $f^{2} - f \not= 0$):
\begin{equation}
\tilde \Gamma^{\alpha}_{\epsilon [  \alpha} \tilde
\Gamma^{\epsilon}_{\nu ]  \mu} =0.
\end{equation}
One can find the metric tensor $g_{\mu \nu}$ from the definition
of the metric connection:
\begin{equation}
g^{\mu \alpha} (g_{\alpha \nu , \epsilon} + g_{\alpha \epsilon ,
\nu} + g_{\nu \epsilon, \alpha})/2 = \gamma^{\mu}_{\nu \epsilon}.
\end{equation} \par
     In a general case, a situation is more complex. For
$ T_{\mu \nu} \not= 0,$ insteed
of  (9) we have the algebraic equations for the manifold $\Sigma^{4}:$
\begin{equation}
2(f^{2} - f) \tilde \Gamma^{\alpha}_{\epsilon [  \alpha} \tilde
\Gamma^{\epsilon}_{\nu ]  \mu}= k (T_{\mu \nu} - g_{\mu \nu} T/2),
\end{equation}
therewith $T_{\mu \nu}$ must be computed on $\Sigma^{4},$ and $g_{\mu
\nu}$ should satisfy the equation (10). \par
     The equation of a trial body motion on $\Sigma^{4}$ on the
coordinates $B^{\mu}:$
\begin{equation}
d^{2}B^{\mu}/ds^{2} +  \gamma^{\mu}_{\nu \epsilon} u^{\nu}u^{\epsilon} = 0,
\end{equation}
where $u^{\nu} = dB^{\nu}/ds,$ can be rewritten on the coordinates
$A^{\tilde \mu}$ as
\begin{equation}
d^{2}A^{\tilde \mu}/ds^{2} + f \tilde \Gamma^{\tilde \mu}_{\tilde \nu
\tilde \epsilon} u^{\tilde \nu}u^{\tilde \epsilon} = 0,
\end{equation}
where $u^{\tilde \nu} = dA^{\tilde \nu}/ds,$ therewith it has such
the view independently from a choice of the coordinates $B^{\mu}.$
A motion of a picked out observation point on $\Sigma^{4}$ is described
by the equation:
\begin{equation}
d^{2}B^{\mu}/ds^{2} + f \Gamma^{\mu}_{\nu \epsilon} u^{\nu}u^{\epsilon} = 0.
\end{equation}

\section[5]{The additional coordinates as fields in the Minkowski space}

\par
     On the first view, the equations (9) are seemed to depend on coordinates
$B^{\mu}.$ But these equations  are easy transformed to the following form:
\begin{equation}
h^{s}_{\tilde \nu}h^{m}_{\tilde \mu}h^{\tilde \mu}_{s [ m}h^{\tilde \nu}_
{c ] d} = 0,
\end{equation}
for which its independence from $B^{\mu}$ is obvious.
The equations (15), rewritten in details, can be interpreted as the
nonlinear differential equations of the
second order for the "field" $A^{\tilde \mu}$ in the Minkowski space $(x):$
\begin{equation}
{{\partial x^{s}} \over {\partial A^{\tilde \nu}}} {{\partial x^{m}} \over
{\partial A^{\tilde
\mu}}} {{\partial^{2} A^{\tilde \mu}} \over {\partial x^{s} \partial
x^{[ m}}}
{{\partial^{2} A^{\tilde \nu}} \over {\partial x^{l ]} \partial x^{d}}} = 0.
\end{equation}
Its coefficients are independent from $B^{\mu},$ therefore $\Sigma^{4}$
is a "cylindrical" hypersurface in the 12-space  $(x,A,B).$

\section[6]{The global symmetries of the model}

\par
     Let us denote $F=f^{2}-f,$ and let $F_{1}=F_{2}$ be the
function values for two values $f_{1}$ and $f_{2}$  of the parameter $f. $ It
follows from Eqs. (7), (8), and (11), that the same connection  $ \gamma^
{\mu}_{\nu \epsilon} $ and two different connections $ \tilde \Gamma^
{\mu}_{\nu \epsilon} $ (so as $\gamma = f \tilde \Gamma + \bar \Gamma)$
correspond to these two values $f_{1}$ and $f_{2}.$ The discrete $D_{1}-
$symmetry will take place on $\Sigma^{4}.$ By an introduction of spinor
fields of the composite fermions \cite{I}, the last symmetry would be
transformed into the global $SU(2)$-symmetry of the unified model.  \par
     The parameter $f$ can have any value, excluding $f=0; 1.$ On the
manifold $\Sigma^{4},$ the global variations of $f$ are not observable.
An existence of two peculiar points will be not essential by localization
of variations of $f.$ So one can consider $U(1)$ to be the global
symmetry group of the model, with the made note. \par
     In a general case, transformations of the groups $SU(2)$ and $U(1)$ will
be connected between themselves.  $SU(2)$-transformations correspond to
"rotations" around the axis $F,$ where $F=f^{2} - f,$ in a transformation
parameters space. It means that a variation of the parameter $f$ can lead
to a permutation of a pair of solutions which will be transformed by
the group $SU(2).$ One needs of additional restrictions to have the
$SU(2) \times U(1)-$symmetry of the model. \par
     The transformations can be independent if:
1) the $SU(2)$-doublets are massless, and 2) a region of permissible
variations of the parameter $f$ is such a one that for any pair $f_{1},
f_{2}:  F(f_{1}) \not= F(f_{2}).$ To satisfy the second condition, one can
perform a rotation in the plane $(f,F)$ on some angle $\theta.$
Under an additional condition that one component of the $SU(2)$-doublet
should be massless after breaking of the $SU(2)$-symmetry
( that is equivalent to the demand $f \to 1$ for the component),
$\theta$ has the minimum value $\theta _{min},$ for which
$\sin^{2}\theta _{min} =0,20$  $(\theta _{min}$ is the angle between
the axis $f$ and the straight line, which goes through the points $(1/2,
-1/4)$  and $(1, 0)$ on the plane $(f, F) ).$ It is approximate enough to
the value of the same function for the Weinberg angle of the standard model.
There is a very close analogy with the situation in the standard model, when
gauge fields of the groups $U(1)$ and $SU(2)$ are linear
transformed to get a gauge field of the observable $U(1)-$symmetry \cite{C}.
\par
     Out of the massless limit, the $SU(2)-$symmetry will be broken
automatically, and the $U(1)-$one will be preserved.

\section[7]{Conclusion}

\par
     The considered linear deformation of the connection, with the global
parameter $f,$ will be a universal method  to embed the general relativity
4-space into a flat 12-space, if any its Ricci tensor would be presented
as the quadratic form (8). This question needs an additional research.
\par
     A local action of the group $U(1)$  should be accompanied by a
statistical description of trajectories of a trial particle, because
non-observable variations of $f$ from one point of the space $(x)$ to
another will provide fluctuations of  particle's trajectory in the space.
\par
    To unify the described model of gravitation in the flat 12-space with
the composite fermions model with the $SU(3)_{c} \times SU(2)_{L}-$symmetry
in the 8-space $(x,y)$ \cite{I}, we would use the flat 16-space
$(x, A=(x_{1}+x_{2})/2, B, y),$ where $x_{1}, x_{2}$ are the coordinates in
the flat 8-space with torsion, for which we had in \cite{I}:
$(x_{1}+x_{2})/2=x, \; (x_{1}-x_{2})/2=y.$ The structure of the discrete
space $(y)$  is caused by symmetry properties of the equations of motion of
the fundamental fermions \cite{I}. Namely the structure leads to an
appearance of the exact $SU(3)_{c}-$symmetry of the composite fermions.
\par
   Equations (9) and (11) are uniform relatively to $\tilde \Gamma.$ It is
an additional advantage, which gives us a possibility to linearize these
equations  relatively to $\tilde \Gamma,$ introducing new variables
\cite{F,Iv}.

\end{document}